\documentclass[english]{article}
\usepackage[T1]{fontenc}
\usepackage[latin9]{inputenc}
\usepackage[letterpaper]{geometry}
\usepackage{array}
\usepackage{textcomp}
\usepackage{graphicx}
\usepackage{fix-cm}
\makeatletter

\@ifundefined{definecolor}
 {\usepackage{color}}{}
\makeatother

\makeatother

\usepackage{babel}
\begin{document}
\title{Comment on "Teleportation of Three-Qubit State via Six-qubit
Cluster State"}
\maketitle
\begin{center}
\author{Anindita Banerjee $^{a}$ \footnote{email: aninditabanerjee.physics@gmail.com} \and Chitra Shukla $^{b}$}

\par\end{center}

\begin{center}
$^{a}$Department of Physics and Center for Astroparticle Physics
and Space Science,
\par\end{center}

\begin{center}
Bose Institute, Block EN, Sector V, Kolkata 700091, India.
\par\end{center}
\begin{center}
$^{b}$Department of Physics and Materials Science and Engineering,
\par\end{center}

\begin{center}
Jaypee Institute of Information Technology, A-10, Sector-62, Noida,
UP-201307, India.
\par\end{center}

\maketitle



\begin{abstract}
Recently Yu  and Sun  {[}Int. J. Theor. Phys. DOI 10.1007/s10773-014-2360-x{]}
have presented probabilistic teleportation of $3$-qubit cat state
via $6$-qubit cluster state. The success probability depends on absolute
value of only two of the  coefficients of  cluster
state i.e. $|c|^{2}+|d|^{2}$. We have demonstrated the feasibility to teleport $3$-qubit
cat state via $2$-qubit non maximally entangled Bell state instead
of a complex state like $6$-qubit non maximally entangled cluster
state. In this comment we have prescribed an optimal protocol for  teleportation of $n$-qubit
state of the form $\left(\alpha|x\rangle+\beta|\bar{x\rangle}\right)_{n}$
 via non maximally entangled Bell state $a|00\rangle+b|11\rangle$
 henceforth allowing teleportation of $n$-qubit state  via
$2$-qubit Bell state. The success probability
of the prescribed protocol is  $|b|^{2}$.

\end{abstract}

\section*{}

Yu and Sun  \cite{Li} have recently presented a scheme
of probabilistic teleportation where they have teleported $3$-qubit
cat state $x|000\rangle+y|111\rangle$ via $6$-qubit non maximally
entangled cluster state $a|000000\rangle+b|000111\rangle+c|111000\rangle+d|111111\rangle$
as the quantum channel. Pathak and  Banerjee \cite{IJQI} have
reported that an $n$-qubit state $\left(\alpha|x\rangle+\beta|\bar{x\rangle}\right)_{n}$
where $x\in\{0,1\}$ can be deterministically teleported via
a Bell state. We can therefore teleport an $n$-qubit state with certain
probability via a $2$-qubit non maximally entangled Bell state.

\begin{figure}
\begin{centering}
\includegraphics[scale=0.4]{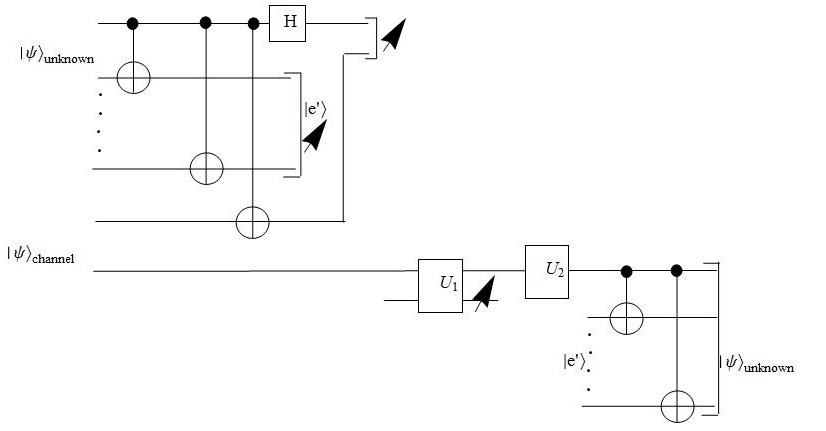}\protect\caption{Quantum circuit for probabilistic teleportation of unknown n-qubit state
$\left(\alpha|x\rangle+\beta|\bar{x\rangle}\right)_{n}$  via non maximally entangled
Bell state.\label{fig:Quantum-circuit}}
\par\end{centering}

\end{figure}

Alice and Bob are spatially separated parties. Alice wish to teleport
an unknown $n$-qubit state in Z-basis of the form as presented in equation (\ref{eq:1}). The quantum circuit of the proposed protocol is also presented in Figure \ref{fig:Quantum-circuit}.

\begin{equation}
\left|\psi_{\textrm{unknown}}\right\rangle =\alpha|x_{1}x_{2}.......x_{n}\rangle+\beta|\bar{x_{1}}\bar{x_{2}}.....\bar{x_{n}}\rangle,\label{eq:1}
\end{equation}  where $x\in \{0,1\}$ and $\alpha$ and $\beta$ are real such that $|\alpha|^{2}+|\beta|^{2}=1$
and $|\alpha|<|\beta|.$  Let all the particles from $2^{nd}$ to the $n^{th}$  be denoted
by $|e\rangle$ and only the first qubit be denoted by $x$. Therefore, $|e\rangle=|x_{2}.......x_{n}\rangle$ and $|\bar{e}\rangle=|\bar{x_{2}}.....\bar{x_{n}}\rangle$. Finally, we can write the unknown $n$-qubit as
\begin{equation}
\left|\psi_{\textrm{unknown}}\right\rangle =\alpha|xe\rangle_{n}+\beta|\bar{x}\bar{e}\rangle_{n}.\label{eq:2}
\end{equation} Alice and Bob possess prior shared non maximally entangled Bell state \begin{equation}
\left|\psi_{\textrm{channel}}\right\rangle =a|00\rangle+b|11\rangle, \label{eq:3}
\end{equation}  where $\left|a\right|^{2}+\left|b\right|^{2}=1$. The final state of the system is given by
\begin{equation}
\begin{array}{lcl}
\left|\psi_{1}\right\rangle& = &\left|\psi_{\textrm{unknown}}\right\rangle \otimes\left|\psi_{\textrm{channel}}\right\rangle\\
 & = & \left(\alpha|xe\rangle+\beta|\bar{x}\bar{e}\rangle\right)\otimes\left(a|00\rangle+b|11\rangle\right)\\
 & = & \alpha a|xe00\rangle+\alpha b|xe11\rangle+\beta a|\bar{x}\bar{e}00\rangle+\beta b|\bar{x}\bar{e}11\rangle.
\end{array}\label{eq:3a}
\end{equation} Alice applies controlled-NOT (CNOT)%
\footnote{CNOT is a two qubit gate, it operates on two qubits $\left|x\right\rangle $and
$\left|y\right\rangle $ such that the output is given by $\left|x\right\rangle $ and
$\left|x\right\rangle \oplus\left|y\right\rangle $ respectively.%
} operations on all her qubits keeping the first qubit as control.
The transformed state is denoted by
\begin{equation}
\begin{array}{ccc}
\left|\psi_{2}\right\rangle  & = & \alpha a|x(x\oplus e)(x\oplus0)0\rangle+\alpha b|x(x\oplus e)(x\oplus1)1\rangle\\
 &  & +\beta a|\bar{x}(\bar{x}\oplus\bar{e})(\bar{x}\oplus0)0\rangle+\beta b|\bar{x}(\bar{x}\oplus\bar{e})(\bar{x}\oplus1)1\rangle.
\end{array}\label{eq:3b}
\end{equation} We will use the following identities in equation (\ref{eq:3b}).
\begin{center}
$\begin{array}{ccc}
x\oplus0 & = & x\\
x\oplus1 & = & \bar{x}\\
\bar{x}\oplus0 & = & \bar{x}\\
\bar{x}\oplus1 & = & x\\
\bar{x}\oplus\bar{y} & = & x\oplus y
\end{array}$
\end{center}
Therefore, after CNOT operations the final state of principal system
is given by

\begin{equation}
\left|\psi_{2}\right\rangle =\alpha a|x(x\oplus e)x0\rangle+\alpha b|x(x\oplus e)\bar{x}1\rangle+\beta a|\bar{x}(x\oplus e)\bar{x}0\rangle+\beta b|\bar{x}(x\oplus e)x1\rangle.\label{eq:4}
\end{equation}

Alice will apply Hadamard gate on her first qubit. Hadamard operation
can be defined as $H|a\rangle=\frac{(-1)^{a}|a\rangle+|\bar{a}\rangle}{\sqrt{2}}$, where $a\in\{0,1\}$. After Hadamard operation the final state
of the system is given by

\begin{equation}
\begin{array}{lcl}
\left|\psi_{3}\right\rangle  & = & \frac{1}{\sqrt{2}}\left((-1)^{x}\alpha a|xe'x0\rangle+\alpha a|\bar{x}e'x0\rangle+(-1)^{x}\alpha b|xe'\bar{x}1\rangle\right.\\
 &  & +\alpha b|\bar{x}e'\bar{x}1\rangle+(-1)^{\bar{x}}\beta a|\bar{x}e'\bar{x}0\rangle+\beta a|xe'\bar{x}0\rangle\\
 &  & \left.+(-1)^{\bar{x}}\beta b|\bar{x}e'x1\rangle+\beta b|xe'x1\rangle\right)
\end{array}\label{eq:7-1}
\end{equation} where $x\oplus e= e'$. In equation (\ref{eq:7-1}) the $|e'\rangle$ is now separable. Thus,
Alice measures her $|e'\rangle$ qubits in computational basis and
communicates them to Bob. Bob will locally prepare $|e'\rangle$ in
his lab. The state of the system is given by  \begin{equation}
\begin{array}{ccc}
\left|\psi_{4}\right\rangle  & = & \frac{1}{\sqrt{2}}\left((-1)^{x}\alpha a|xx0\rangle+\alpha a|\bar{x}x0\rangle+(-1)^{x}\alpha b|x\bar{x}1\rangle+\alpha b|\bar{x}\bar{x}1\rangle+\right.\\
 &  & \left.(-1)^{\bar{x}}\beta a|\bar{x}\bar{x}0\rangle+\beta a|x\bar{x}0\rangle+(-1)^{\bar{x}}\beta b|\bar{x}x1\rangle+\beta b|xx1\rangle\right).
\end{array}\label{eq:6}
\end{equation} The equation (\ref{eq:6}) resembles the teleportation of unknown qubit
$\alpha|x\rangle+\beta|\bar{x}\rangle$ when channel is maximally
entangled. Therefore, we are basically teleporting one qubit and rest
of the qubits $(n-1)$ are classically communicated. Alice measures
her qubits again in computational basis and communicates to Bob. Bob will apply a suitable operator $U_{2}$  where $U_{2}\in{I, X, Z,  iY }$ on his qubit.

Bob does not want to do extra work so before working on the qubit
he wants to be sure if the  protocol fails or succeeds.
If the scheme succeeds then only he will proceed. If Alice's measurement
result is $xx$, then resultant state gets reduced to $\frac{(-1)^{x}\alpha a|0\rangle+\beta b|1\rangle}{\sqrt{\left|\alpha a\right|^{2}+\left|\beta b\right|^{2}}}$. Bob creates an ancilla $|0\rangle$ and applies a unitary $U_{1}$
on the combined system of his qubit and  ancilla
where\\
\begin{center}
$U_{1}=\left(\begin{array}{cccc}
\frac{b}{a} & \sqrt{1-\frac{b^{2}}{a^{2}}} & 0 & 0\\
0 & 0 & 0 & -1\\
0 & 0 & 1 & 0\\
\sqrt{1-\frac{b^{2}}{a^{2}}} & -\frac{b}{a} & 0 & 0
\end{array}\right).$
\end{center} Final state of the system is \begin{equation}
\left|\psi_{5}\right\rangle =\frac{1}{\sqrt{|\alpha a|^{2}+|\beta b|^{2}}}\left[b(\alpha|0\rangle+\beta|1\rangle)|0\rangle+\sqrt{1-\frac{b^{2}}{a^{2}}}\alpha a|1\rangle|1\rangle\right]\label{eq:6a}.
\end{equation} Bob measures the ancilla in the computational basis. If he gets $|0\rangle$,
then the protocol is successful and if he gets $|1\rangle$ then the
protocol fails. Iff, he gets $|0\rangle$ then with $b^{2}$ probability he has obtained
the state

\[
\left|\psi_{6}\right\rangle =\alpha|0\rangle+\beta|1\rangle.
\]  Bob will apply $U_{2}$ on his shared qubit and obtain the unknown
qubit which in this case is Identity operator. For other results of
Alice $(01/10/11)$, we refer to the standard teleportation protocol
for which Bob needs to apply suitable operator$(X/Z/iY)$ . Now, he
has to obtain all the $n$-qubits in his lab thus, he will produce
the separable bits $|e'\rangle^{\otimes n-1}$. Thus, the resultant state is
\begin{equation}
\begin{array}{lcl}
\left|\psi_{7}\right\rangle  & = & \alpha|0e'\rangle+\beta|1e'\rangle\\
 & = & \alpha|0(x\oplus x_{2},.......,x\oplus x_{n})\rangle+\beta|1(x\oplus x_{2},.......,x\oplus x_{n})\rangle.
\end{array}\label{eq:7}
\end{equation} Bob applies CNOT operations on $|e'\rangle^{\otimes n-1}$  and obtains the  $n$-qubit state as shown in equation (\ref{eq:8}).

\begin{equation}
\begin{array}{ccc}
\left|\psi_{8}\right\rangle  & = & \alpha|x_{1}x_{2...}x_{n})\rangle+\beta|\bar{x_{1}}\bar{x_{2}}.......\bar{x_{n}})\rangle\end{array}\label{eq:8}
\end{equation}

Therefore in this letter we have shown that an  $n$-qubit state $\left(\alpha|x\rangle+\beta|\bar{x\rangle}\right)_{n}$
where $x\in\{0,1\}$ can be teleported via a $2$-qubit Bell state. This can be done with  success probability  $b^{2}$.

\section*{}

\textbf{Acknowledgment:}Authors thank Anirban Pathak and Som Shubhro Bandyopadhyay for discussion. AB thanks  DST-SERB project
SR/S2/LOP-18/2012.

\end{document}